\documentclass[twocolumn]{aastex63}
\usepackage{bm}

\newcommand{\AP}{\color{black}}

\date{\today}
\submitjournal{AJ}

\shorttitle{Non-Boussinesq low-Prandtl number convection}
\shortauthors{A. Pandey, J. Schumacher, \& K. R. Sreenivasan}

\begin{document}

\title{Non-Boussinesq low-Prandtl number convection with a temperature-dependent thermal diffusivity}

\correspondingauthor{Ambrish Pandey}
\email{ambrish.pandey@nyu.edu}

\author[0000-0001-8232-6626]{Ambrish Pandey}
\affiliation{Center for Space Science, New York University Abu Dhabi, Abu Dhabi 129188, United Arab Emirates}

\author[0000-0002-1359-4536]{J\"org Schumacher}
\affiliation{Institut f\"ur Thermo- und Fluiddynamik, Technische Universit\"at Ilmenau, Postfach 100565, D-98684 Ilmenau, Germany}
\affiliation{Tandon School of Engineering, New York University, New York 11201, USA}

\author{Katepalli R. Sreenivasan}
\affiliation{Center for Space Science, New York University Abu Dhabi, Abu Dhabi 129188, United Arab Emirates}
\affiliation{Tandon School of Engineering, New York University, New York 11201, USA}

\begin{abstract}
In an attempt to understand the role of the strong radial dependence of thermal diffusivity on the properties of convection in sun-like stars, we mimic that effect in non-Oberbeck-Boussinesq (NOB) convection in a horizontally-extended rectangular domain (aspect ratio 16), by allowing the thermal diffusivity $\kappa$ to increase with the temperature (as in the case of stars). Direct numerical simulations (i.e., numerical solutions of the governing equations by resolving up to the smallest scales without requiring any modeling) show that, in comparison with Oberbeck-Boussinesq (OB) simulations (two of which we perform for comparison purposes), the symmetry of the temperature field about the mid-horizontal plane is broken, whereas the velocity and heat flux profiles remain essentially symmetric. Our choice of $\kappa(T)$, which resembles the variation in stars, results in the temperature field that loses its fine structures towards the hotter part of the computational domain, but the characteristic large scale of the turbulent thermal `superstructures', which are structures whose size is typically larger than the depth of the convection domain, continue to be largely independent of the depth. 
\end{abstract}


%%%%%%%%%%%%%%%%%%%%%%================%%%%%%%%%%%%%%%%%%%%%%%%%%%%%
\section{Introduction}

Thermal convection is the primary driving mechanism in many turbulent flows in nature. For instance, it is responsible for the transport of heat and momentum in the convection zone of the Sun, sun-like stars and many of their planets~\citep{Miesch:LRSP2005, Nordlung:LRSP2009, Rincon:LRSP2018, Hanasoge:ARFM2016, Schumacher:RMP2020}. The Rayleigh-B\'enard convection (RBC), which is a paradigm of thermal convection systems, consists of a fluid layer confined between two horizontal plates heated from below and cooled from above~\citep{Ahlers:RMP2009, Chilla:EPJE2012, Verma:book2018}. The main governing parameters of RBC are the Rayleigh and Prandtl numbers. The Rayleigh number $Ra$ signifies the strength of driving force by buoyancy relative to dissipative forces due to viscosity and thermal diffusivity, and the Prandtl number $Pr$ is defined as the ratio of the kinematic viscosity and thermal diffusivity of the fluid. The aspect ratio $\Gamma$, which is the ratio of the horizontal to the vertical dimensions of the flow domain, also influences the flow. Flow domain in natural convective systems are usually horizontally extended, which leads to the formation of regular large-scale patterns which are denoted as {\em turbulent superstructures}. These are, for example, granules and supergranules in the Sun~\citep{Nordlung:LRSP2009} and other stars~\citep{Michel:Science2008} or cloud streets in the atmosphere~\citep{Markson:Science1975}. Moreover, convection in the Sun and other stellar objects are characterized by low $Pr$ and very high $Ra$~\citep{Nordlung:LRSP2009, Hanasoge:ARFM2016, Schumacher:RMP2020}, accompanied by strong spatial variations of fluid properties such as pressure, density and temperature. 

Convection in low-$Pr$ fluids is dominated by inertia and the flow is highly turbulent compared to those for unity-$Pr$ fluids~\citep{Schumacher:PNAS2015, Schumacher:PRF2016, Scheel:JFM2016, Pandey:Nature2018, Schumacher:RMP2020}. Direct numerical simulations (DNS) of very low-$Pr$ turbulent convection at high-$Ra$ without parametrizations of turbulence are challenging due to very small length and time scales, which need to be resolved properly in order to faithfully study such flows. Large $\Gamma$ places even greater demands on resources but this has been achieved recently~\citep{Pandey:Nature2018}. However, these simulations were made under the Oberbeck-Boussinesq (OB) approximation, where the density variation is neglected in all terms of the governing equations except in the buoyancy term. Additionally, molecular transport coefficients, e.g., thermal diffusivity, kinematic viscosity, and thermal expansion coefficient were assumed to be constant and independent of temperature. 

The OB approximation is obviously not applicable to the stellar environment which is characterized by a very strong stratification, which is in line with orders-of-magnitude variations of the material parameters and state variables such as temperature, pressure and mass density across the depth of the convection zone~\citep{Dalsgaard:Science1996, Hanasoge:ARFM2016, Schumacher:RMP2020}. {\AP For the Sun, the thermal conductivity due to photons displays a cubic dependence on temperature, $k(T)\sim T^3$, thus varying more strongly than the kinematic viscosity due to ions, $\nu(T) \sim T^{5/2} \rho^{-1}$, which causes the molecular Prandtl number to  be of the order $10^{-6}$ throughout the convection zone strongly even more towards the surface \citep{Miesch:LRSP2005, Freytag:JCP2012, Schumacher:RMP2020}. Therefore, we think that the effect of varying thermal conductivity $k(T)$ and thus of the thermal diffusivity $\kappa(T)$ is the more dominant effect to include for purposes of generating in our simple model with constant $\nu$ the right qualitative trend of $Pr$ across the convection zone. We know from preliminary calculations that the additional inclusion of $\nu(T)$ breaks the symmetry of the velocity field and heat flux, which is not observed by including only $\kappa(T)$. These results will be reported separately as the next logical step.}

Specifically, we perform direct numerical simulations of convection in an extended rectangular domain (aspect ratio 16) by letting the thermal diffusivity in the governing equations to depend on the  temperature. This model introduces a strong temperature stratification towards the top of the domain, resulting in a depth-variation of $Pr$ similar to that in the Sun \citep{Schumacher:RMP2020}. To serve as a reference, we perform two simulations of RBC in the Boussinesq limit with Prandtl and Rayleigh numbers corresponding to those at the top and bottom plates of our NOB convection run. Our major finding is that some of the flow properties in the NOB convection, such as the profiles of velocity and heat transport are largely unaffected, whereas the properties of the temperature field are different from those in the OB convection. Further, the variation of the small-scale content of superstructures in the temperature field decreases with depth, although the underlying large-scale structure remains essentially unchanged.

Properties of turbulent superstructures, such as their characteristic spatial and temporal scales and their contribution in the global heat transport, have been studied for a range of Prandtl and Rayleigh numbers~\citep{Pandey:Nature2018, Schneide:PRF2018, Fonda:PNAS2019, Stevens:PRF2018, Green:JFM2020, Krug:JFM2020}. These numerical studies, however, adopted the OB model of convection, which, as we just implied, does not accurately yield the properties of turbulent superstructures in stellar convection. It is therefore important to systematically explore the non-Oberbeck-Boussinesq effects on the properties of turbulent superstructures of convection. This is one motivation of the current work. Another motivation is to study the statistical nature of the velocity and temperature fields.

In the past, NOB convection has been explored primarily to study the properties of heat transport and flow structures in confined flow domains of aspect ratio $\Gamma \approx 1$~\citep{Wu:PRA1991, Zhang:POF1997, Ahlers:JFM2006, Sameen:PhSc2008, Sameen:EPL2009, Horn:JFM2013, Valori:PRE2017, Valori:EiF2019}. NOB effects manifest, among other effects, by a mean temperature that is different from the arithmetic mean of top and bottom temperatures, which leads to symmetry breaking characteristic of OB convection~\citep{Hurlburt:APJ1984, Wu:PRA1991, Zhang:POF1997, Evonuk:GAFD2004, Skrbek:JFM2015}. In NOB convection, the width of the thermal and viscous boundary layers (BLs) as well as the temperature drop within the thermal BLs at the two plates are different~\citep{Wu:PRA1991, Ahlers:JFM2006, Horn:JFM2013}. In \cite{Sameen:EPL2009}, the NOB effects were systematically explored by varying various temperature-dependent fluid properties one at a time and keeping others temperature-independent. Recently, \cite{Getling:PLA2018} incorporated temperature-dependent thermal diffusivity in their DNS in a horizontally-extended domain to study the multiscale nature of the solar convective flows. They observed that the flow near the isothermal boundary is a superposition of cellular structures with different characteristic scales, which qualitatively agree with those observed in the Sun~\citep{Getling:PLA2018}.

%%%%%%%%%%%%%%%%%%%%%%================%%%%%%%%%%%%%%%%%%%%%%%%%%%%%
\section{Governing equations and formalism}	\label{sec:formalism}

In this work, we extend the OB model of thermal convection by considering only $\kappa$ as a function of temperature, while neglecting the temperature dependence of molecular viscosity and the thermal expansion coefficient. {\AP We regard this approach as a first step in systematically understanding the NOB effects in an extended flow domain that is relevant to natural convective flows. A considerable effort has already been devoted to the study of the solar and astrophysical convection by simulating the governing equations in fully compressible and anelastic limits, along with other driving factors such as Coriolis and Lorentz forces, differential rotation, etc. Global simulations in the anelastic limit of solar convection have been reported for example by~\cite{Miesch:APJ2008, Kapyla:AR2010, Nelson:SP2014, Hotta:Science2016} or \cite{Guerrero:APJ2016}. Simulations of the upper convection zone, which are fully compressible and include detailed radiative transfer, were conducted by \cite{Stein:APJ1998, Gudiksen:AA2011, Freytag:JCP2012, Magic:AA2013, Riethmuller:AA2014} or \cite{Wray:arXiv2015}. All these simulation models have to apply subgrid models for the unresolved scales, rely on numerical diffusion, or are even run as implicit (large-eddy) simulations. Our long-term goal, however, is to systematically and incrementally include the complexities present in realistic astrophysical flows in direct numerical simulation models starting with their influence in the simplest case of RBC. They will provide helpful insights on scaling of effective eddy viscosity and diffusivity that have to be used in full-size simulations.}

We solve the following governing equations
\begin{eqnarray}
\frac{\partial {\bm u}}{\partial t} + {\bm u} \cdot \nabla {\bm u} & = & -\frac{\nabla p}{\rho_0} + \gamma g T \hat{z} + \nu \nabla^2 {\bm u}, \label{eq:u} \\
\frac{\partial T}{\partial t} + {\bm u} \cdot \nabla T & = & \nabla \cdot [ \kappa(T) \nabla T],
 \label{eq:T} \\
\nabla \cdot {\bm u} & = & 0 \label{eq:m},
\end{eqnarray}
where ${\bm u} \, ( = u_x \hat{x} + u_y \hat{y} + u_z \hat{z})$, $T$, and $p$ are the velocity, temperature and pressure fields, respectively. The coefficients $\gamma$ and $\nu$ are the thermal expansion coefficient and the kinematic viscosity of the fluid, respectively, {\AP $\rho_0$ is the mean density of the fluid}, and $g$ is the acceleration due to gravity. Note that the last term in the temperature equation [Eq.~(\ref{eq:T})] becomes $\kappa \nabla^2 T$ in OB convection.

The temperature drops very quickly over a relatively short distance near the photosphere in the convection zone of the Sun~\citep{Schumacher:RMP2020}, so we choose a particular form of $\kappa(T)$ to mimic such a temperature drop in our flow domain near the top plate~\citep{Getling:PLA2018}. We consider $\kappa(T)$ as a polynomial in $T$ such that
\begin{equation}
\kappa(T) = \kappa_{\mathrm{top}} f(T) \label{eq:kappa}
\end{equation}
with
\begin{equation}
f(T) =  1 + a T^{\alpha} + b T^{\beta}, \label{eq:f}
\end{equation}
where $\kappa_{\mathrm{top}}$ is thermal diffusivity at the top plate. It is clear that, for positive $a, b, \alpha, \beta$, the thermal diffusivity increases with depth due to increasing temperature. As a result (see ~\cite{Busse:JFM1967}), the conduction temperature profile does not remain linear as in OB convection (see Figure~\ref{fig:conduction}). In the following, we will describe the consequences of temperature-dependent $\kappa$ on the velocity and temperature profiles, as well as on the relations between global heat transport and dissipation rates~\citep{Howard:ARFM1972, Shraiman:PRA1990}.

%%%%%%%%%%%%%%%%%%%%%%=========%%%%%%%%%%
\subsection{Conduction temperature profile}	\label{subsec:Tc}
In the conduction state of no fluid motion, temperature varies only in the vertical direction, i.e., $\bm u =  0$ and $T = T_c(z)$. Therefore, Eq.~(\ref{eq:T}) in the conduction state becomes
\begin{equation}
\frac{\partial}{\partial z} \left( \kappa_c (T_c) \frac{\partial T_c}{\partial z} \right) = 0,
\end{equation}
which yields
\begin{equation}
\kappa_c (T_c) \frac{\partial T_c}{\partial z}  = c_1, \label{eq:cond0}
\end{equation}
where $c_1$ is a constant of integration. Integration of this equation and the use of the explicit temperature dependence of $\kappa$ from Eqs.~(\ref{eq:kappa}) and (\ref{eq:f}) gives
\begin{equation}
\kappa_{\mathrm{top}}  \left( T_c + \frac{a}{\alpha+1} T_c^{\alpha+1} + \frac{b}{\beta+1} T_c^{\beta+1} \right)= c_1 z + c_2, \label{eq:cond1}
\end{equation}
where $c_2$ is another constant of integration. The constants $c_1$ and $c_2$ can be determined by specifying the temperature boundary conditions at the top and bottom plates. For the isothermal boundary conditions, i.e., $T(z=0) = T_{\mathrm{bot}} = 1$ and $T(z=1) = T_{\mathrm{top}}=0$, the constants $c_1$ and $c_2$ can be fixed. The boundary condition at $z=0$ gives
\begin{equation}
c_2 = \kappa_{\mathrm{top}} \left( 1 + \frac{a}{\alpha+1} + \frac{b}{\beta+1} \right) =: \kappa_{\mathrm{top}} C,
\end{equation}
with
\begin{equation}
C = 1 + \frac{a}{\alpha+1} + \frac{b}{\beta+1},	\label{eq:C}
\end{equation}
and with the condition at $z=1$, one obtains
\begin{equation}
c_1 = -c_2 = -\kappa_{\mathrm{top}} C\,.
\end{equation}
Using the above $c_1$ and $c_2$ in Eq.~(\ref{eq:cond1}), the conduction temperature profile can be determined by solving
\begin{equation}
T_c(z) + \frac{a}{\alpha+1} T_c^{\alpha+1}(z) + \frac{b}{\beta+1} T_c^{\beta+1}(z) = C(1-z). \label{eq:cond2}
\end{equation}
Note that the shape of resulting nonlinear conduction temperature profile depends on the functional form of $\kappa(T)$, i.e., on the coefficients $a$, $b$, $\alpha$, and $\beta$ in Eq.~(\ref{eq:f}).

%%%%%%%%%%%%%%%%%%%%%%=========%%%%%%%%%%
\subsection{Conductive heat transport}	\label{subsec:Jc}
Heat flux in the conduction state is defined as
\begin{equation}
J_c = -\kappa_c(T_c) \nabla T_c = -\kappa_c(T_c) \frac{\partial T_c}{\partial z}, \label{eq:heat_cond}
\end{equation}
where $\kappa_c(T_c)$ is the thermal diffusivity profile in the conduction state. Note that $J_c$ simply becomes $\kappa \Delta T/H$ in OB convection due to a fixed temperature gradient over the depth of the fluid layer. In the present case, however, the conductive heat flux can be estimated using Eq.~(\ref{eq:cond0}) as
\begin{equation}
J_c = -c_1 = \kappa_{\mathrm{top}} C,
\end{equation}
where $C = C(a, b, \alpha, \beta)$ is given by Eq.~(\ref{eq:C}). Thus $J_c$ can be estimated analytically once the coefficients $a, b, \alpha, \beta,$ and $\kappa_{\mathrm{top}}$ are specified. Moreover, $J_c$ is a constant at each point of the flow, and is therefore the same as the volume-averaged conductive heat flux, i.e.,
\begin{equation}
\langle J_c \rangle_{V} = J_c = \kappa_{\mathrm{top}} C,
\end{equation}
where $\langle \cdot \rangle_V$ is the average over the entire domain.

%%%%%%%%%%%%%%%%%%%%%%=========%%%%%%%%%%
\subsection{Turbulent heat transport}	\label{subsec:flux}
To determine the heat transport in the convective state, we rewrite the temperature equation [Eq.~(\ref{eq:T})] using the mass conservation as
\begin{equation}
\frac{\partial T}{\partial t} + \nabla \cdot ({\bm u} T - \kappa(T) \nabla T) = 0. \label{eq:heat}
\end{equation}
Horizontal-averaging of Eq~(\ref{eq:heat}) yields
\begin{equation}
\frac{\partial \langle T \rangle_A}{\partial t} + \langle  \nabla \cdot ({\bm u} T - \kappa(T) \nabla T) \rangle_A = 0,
\end{equation}
where $\langle \cdot \rangle_A$ denotes a horizontally-averaged quantity. Using the no-slip condition for the velocity field and the adiabatic condition for the temperature field on the sidewalls, and additionally averaging in time, we get
\begin{equation}
\left\langle \frac{\partial}{ \partial z} \left( u_z T - \kappa(T) \frac{\partial T}{ \partial z} \right) \right \rangle_{A,t} = 0, 
\end{equation}
which yields 
\begin{equation}
\langle u_z T \rangle_{A,t} - \left \langle  \kappa(T)  \frac{\partial T}{ \partial z}  \right \rangle_{A,t} = J_z,	\label{eq:J_z}
\end{equation}
where $J_z$ is a constant. Note that the use of periodic conditions on the sidewalls also results in Eq.~(\ref{eq:J_z}).
A non-dimensional measure of the convective heat transport is the Nusselt number $Nu$, defined as the ratio of the total heat transport to the conductive heat transport, which is computed for each horizontal plane as 
\begin{equation}
Nu(z) = \frac{\langle u_z T \rangle_{A,t} - \langle  \kappa(T)  \partial T/ \partial z  \rangle_{A,t}}{J_c}. \label{eq:Nu_z}
\end{equation}
The boundary conditions and the conservation of internal energy implies that $Nu(z)$ remains a constant across the fluid layer in the NOB convection, as in OB convection. Integrating Eq.~(\ref{eq:Nu_z}) in the vertical direction yields the volume-averaged Nusselt number
\begin{equation}
Nu = \frac{\langle u_z T \rangle_{V,t} - \langle  \kappa(T) \partial T/\partial z \rangle_{V,t}}{J_c},
\end{equation}
which says that the total heat flux results from the convective as well as the diffusive heat fluxes. Note that the volume-averaged diffusive heat transport $J_d = - \langle  \kappa(T) \partial T/ \partial z \rangle_{V,t}$ in the NOB convection is not equal to $J_c$, unlike in the OB convection. Thus, the volume-averaged Nusselt number in the NOB convection with $\kappa(T)$ is computed as
\begin{equation}
Nu = \frac{\langle u_z T \rangle_{V,t} + J_d}{J_c}. \label{eq:Nu_V}
\end{equation}

%%%%%%%%%%%%%%%%%%%%%%=========%%%%%%%%%%
\subsection{Exact relations} \label{subsec:exact}
Exact relations relating the volume-averaged dissipation rates to the global heat transport in RBC can be derived by volume- and time-averaging the equations governing the temporal evolutions for the kinetic and thermal energies~\citep{Howard:ARFM1972, Shraiman:PRA1990}. The viscous and thermal dissipation rates at each point in the flow are defined, respectively, as 
\begin{eqnarray}
\varepsilon_u & = & \frac{\nu}{2} \left( \frac{\partial u_i}{\partial x_j} +  \frac{\partial u_j}{\partial x_i} \right)^2, \label{eq:eps_u} \\ 
\varepsilon_T & = & \kappa(T) \left( \frac{\partial T}{\partial x_i} \right)^2, 	\label{eq:eps_T}
\end{eqnarray}
where $u_i$ and $x_i$ are the components of the velocity and position vector. Following the similar steps as in \cite{Howard:ARFM1972, Shraiman:PRA1990}, one can obtain the exact relations in convection with temperature-dependent thermal diffusivity as
\begin{eqnarray}
\langle \varepsilon_u \rangle_{V,t} & = & \gamma g [J_c Nu - J_d], \label{eq:exact_u} \\
\langle \varepsilon_T \rangle_{V,t} & = & J_c \frac{\Delta T}{H} Nu, \label{eq:exact_T}  
\end{eqnarray}
where $\Delta T$ and $H$ are the temperature difference and distance between the top and bottom plates.
In the next section, we describe details of DNS of convection in both OB and NOB cases.

%%%%%%%%%%%%%%%%%%%%%%================%%%%%%%%%%%%%%%%%%%%%%%%%%%%%
\section{Details of direct numerical simulations} 	\label{sec:dns}

We investigate convection flow in a three-dimensional rectangular box of horizontal dimensions $L_x = L_y = 16H$, where the top and bottom plates are kept at (different) constant temperatures and satisfy no-slip boundary condition. Periodic boundary conditions are employed on the sidewalls~\citep{Bekki:APJ2017}. Convective flow properties in horizontally-extended domains with periodic sidewalls have been observed to be nearly the same as those with solid sidewalls~\citep{Pandey:Nature2018}. In any case, periodic boundary conditions rather than solid walls are more relevant to convection in the Sun and other stars. We non-dimensionalize the governing equations using $H$, $\Delta T$, $U_f$, and $T_f$ as the length, temperature, velocity, and time scales, respectively. Here $U_f = \sqrt{\gamma g \Delta T H}$ is the free-fall velocity and $T_f= H/U_f$ is the free-fall time. The resulting non-dimensional governing equations are
\begin{eqnarray}
\frac{\partial {\bm u}}{\partial t} + {\bm u} \cdot \nabla {\bm u} & = & -\nabla p + T \hat{z} + \sqrt{\frac{Pr_{\mathrm{top}}}{Ra_{\mathrm{top}}}} \, \nabla^2 {\bm u}, \label{eq:u_n} \\
\frac{\partial T}{\partial t} + {\bm u} \cdot \nabla T & = & \nabla \cdot \left[ \frac{f(T)}{\sqrt{Pr_{\mathrm{top}} Ra_{\mathrm{top}} }} \nabla T \right],
 \label{eq:T_n} \\
\nabla \cdot {\bm u} & = & 0 \label{eq:m_n},
\end{eqnarray}
where the Rayleigh and Prandtl numbers are specified at the top plate as $Ra_{\mathrm{top}} = \gamma g \Delta T H^3/(\nu \kappa_{\mathrm{top}})$ and $Pr_{\mathrm{top}} = \nu/\kappa_{\mathrm{top}}$, and $f(T)$ is given by Eq.~(\ref{eq:f}). Due to varying $\kappa(T)$ with depth, the Rayleigh and Prandtl numbers, which are defined as $Ra = \gamma g \Delta T H^3/\nu \kappa$ and $Pr = \nu/\kappa$, also vary in the flow. Note that the temperature dependence of $\kappa$ only appears in Eq.~(\ref{eq:T_n}) as the viscosity $\nu$ is kept constant everywhere.

The governing equations (\ref{eq:u_n})--(\ref{eq:m_n}) are solved using an exponentially fast converging spectral element solver {\sc Nek5000}~\citep{Fischer:JCP1997}. As in \cite{Scheel:NJP2013}, the simulation domain is divided into a finite number of elements ($N_e$), and on each spectral element the turbulence fields are expanded with Lagrangian interpolation polynomials of order $N$. For this work, we choose $f(T) = 1+49T+450T^6$, and therefore, the thermal diffusivity varies with temperature as
\begin{equation}
\kappa(T) = \kappa_{\mathrm{top}}(1+49T+450T^6) \, . \label{eq:kappa_tw}	
\end{equation}
% kappa_tw == kappa in this work
Thus, the thermal diffusivity at the bottom plate $\kappa_{\mathrm{bot}} = 500 \kappa_{\mathrm{top}}$. 
We perform a direct numerical simulation using this $\kappa(T)$ for $Pr_{\mathrm{top}} = 12.73$ and 
$Ra_{\mathrm{top}} = 1.708 \times 10^8$, which correspond to $Pr_{\mathrm{bot}} = 0.02546$ and 
$Ra_{\mathrm{bot}} = 3.416 \times 10^5$. Thus, the Rayleigh and Prandtl numbers at the bottom of the flow are 500 times smaller than those at the top. 

We mention here that the critical Rayleigh number, $Ra_\mathrm{crit}$, at which heat cannot be transported by molecular diffusion alone and convection comes into effect, is not the same as $Ra_\mathrm{crit} = 1708$ for the OB convection~\citep{Chandrasekhar:Book}. We determined the critical Rayleigh number, $Ra_\mathrm{crit}$, at which temperature profile starts to depart from the conduction profile $T_c(z)$ (shown in Figure~\ref{fig:conduction})~\citep{Getling:PLA2018}. For $\kappa(T) = \kappa_{\mathrm{top}}(1+49T+450T^6)$, we find $Ra_\mathrm{crit} \approx 7 \times 10^5$, more than two orders of magnitude higher than the $Ra_\mathrm{crit}$ for OB convection. {\AP Note that critical $Ra$ also increases in compressible convection with increasing density stratification~\citep{Hurlburt:APJ1984, Gastine:Icarus2012}.}

For comparison purposes, we also perform two OB convection simulations with constant $\kappa$. One simulation has the Rayleigh and Prandtl numbers corresponding to the top plate of our NOB simulation, i.e., for $Pr = 12.73, Ra = 1.708 \times 10^8$; another has $Ra$ and $Pr$ corresponding to those at the bottom plate, i.e, for $Pr = 0.02546, Ra = 3.416 \times 10^5$. These two OB simulations will be referred to hereafter as $\mathrm{OB_{top}}$ and $\mathrm{OB_{bot}}$, respectively (or jointly as OB simulations), whereas the NOB convection simulation with the temperature-dependent thermal diffusivity will be referred to as NOB simulation. The important details of the DNS are summarized in Table~\ref{table:details}. Some further comments on the structure of the computational grid, imposed by the asymmetry of the temperature profiles, can be found in the next section.
%%%%%%%%%%%%%%%%%%%%%%%%%%%%%%%%%%%
\begin{table*}[ht!]
\begin{ruledtabular}
\caption{Important parameters of DNS in a three-dimensional (3D) rectangular box of dimensions $L_x:L_y:H = 16:16:1$; $N_e$ is the total number of elements in the entire flow domain and $N$ is the order of Lagrangian polynomials for the spectral interpolation within each element. For OB simulations, the flow domain is divided into $220^2 \times 32 = 1548800$ elements. For the NOB simulation with temperature-dependent thermal diffusivity, the flow domain is divided into $320^2 \times 87 = 8908800$ elements, where the width and length of the elements are the same everywhere but the height of the elements varies with the depth; as discussed in Section~\ref{subsec:profiles}, the density of elements in the vicinity of the top plate is much larger than near the bottom. The $Pr$ and $Ra$ values for the NOB case vary with the depth, and their values at the top and bottom plates are the same as those of the OB simulations, OB$_{\mathrm{top}}$ and OB$_{\mathrm{bot}}$. $Nu, Re$, and $u_\mathrm{rms}$, are the volume- and time-averaged Nusselt number, Reynolds number, and the root-mean-square (rms) velocity, respectively. $\lambda_U, \lambda_\Theta$, and $\tau$ are the characteristic spatial and temporal scales of turbulent superstructures~\citep{Pandey:Nature2018}. $t_\mathrm{total}$ is the total integration time in units of the free-fall time (in reference to the top wall conditions for the NOB case).
}
\begin{tabular}{ccccccccccccc}
Run & $\kappa/\kappa_{\mathrm{top}}$ & $Pr$ & $Ra$ & $N_e$ & $N$ & $Nu$ & $u_{\mathrm{rms}}$ & $Re$ & $\lambda_U/H$ & $\lambda_\Theta/H$ & $\tau \, (T_f)$ & $t_{\rm total} \, (T_f) $ \\
\hline
NOB & $1+49T+450T^6$ & variable & variable & 8908800 & 5 &  5.4 & 0.300 & 1100 & 4.0 & 4.0 & 47 & 24 \\
$\mathrm{OB_{top}}$ & 1 & 12.73 & $1.708 \times 10^8$ & 1548800 & 9 & 35.8 & 0.094 & 345 & 4.7 & 8.0 & 169 & 128  \\
$\mathrm{OB_{bot}}$ & 1 & 0.025 & $3.416 \times 10^5$ & 1548800 & 9 & 3.65 & 0.469 & 1736 & 4.0 & 4.0 & 30 & 18 \\
\end{tabular}
\label{table:details}
\end{ruledtabular}
\end{table*}
%%%%%%%%%%%%%%%%%%%%%%%%%%%%%%%%%%%

%%%%%%%%%%%%%%%%%%%%%%================%%%%%%%%%%%%%%%%%%%%%%%%%%%%%
\section{Velocity and temperature statistics}	\label{sec:results}

From the numerical data, we compute several quantities to compare the NOB case with OB simulations. We first discuss the depth dependence of horizontally-averaged quantities.

%%%%%%%%%%%%%%%%%%%%%%=========%%%%%%%%%%
\subsection{Vertical profiles}	\label{subsec:profiles} 
\begin{figure}[ht!]
\includegraphics[width=0.47\textwidth]{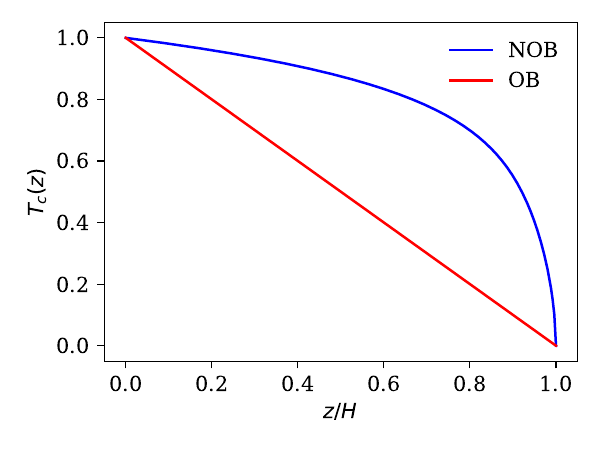}
\caption{Conduction temperature profiles for the OB and NOB cases. It is nonlinear for the latter, with strongly dissimilar structure at the bottom and top plates.}
\label{fig:conduction}
\end{figure}

The temperature profile in the conduction state is linear in OB convection, whereas that in NOB convection becomes nonlinear, as shown in Figure~\ref{fig:conduction}. The temperature gradients are different at the two horizontal plates. Using Eq.~(\ref{eq:cond0}), we can estimate the ratio of the temperature gradients at the two plates in the conduction state as
\begin{equation}
\frac{(\partial T_c/\partial z)_{\mathrm{top}}}{ (\partial T_c/\partial z)_{\mathrm{bot}}} = \frac{ \kappa_{\mathrm{bot}}} {\kappa_{\mathrm{top}}} = 500,
\end{equation}
showing that, in the vicinity of the top plate, the temperature drops very sharply to $T = 0$. Figure~\ref{fig:conduction} shows that $\partial T_c/\partial z$ is negative at all depths, that is, the flow is unstable everywhere. 
%===========================================================
%=========================================================== 

\begin{figure}[ht!]
\includegraphics[width=0.47\textwidth]{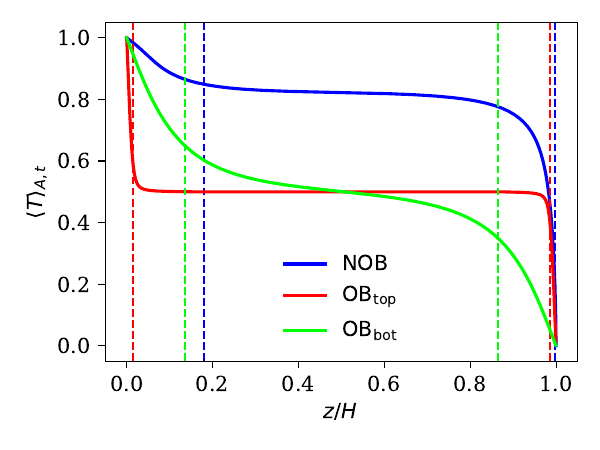}
\caption{Vertical temperature profiles for all the simulations. Averaged temperature in the bulk region for the NOB simulation is approximately 0.82, whereas it is 0.5 for the OB simulations. The thermal BL thicknesses $\delta_T$ are indicated by vertical dashed lines with the same colors as the corresponding temperature profiles. It is clear that $\delta_T$ values for the NOB simulation are quite different at the top and bottom plates, unlike for OB simulations for which they are equal. }
\label{fig:T_z}
\end{figure}

In the convective state shown in Figure~\ref{fig:T_z}, enhanced mixing leads to approximately a constant temperature in the bulk region of the flow. We demonstrate this by plotting the horizontally- and temporally-averaged temperature field $\langle T \rangle_{A,t}$ as a function of the depth $z$. The Figure shows that the averaged temperature in the bulk, i.e., roughly between $z = 0.2$ and $z = 0.8$, does not vary appreciably, and that $T_\mathrm{bulk} \approx 0.82$, which is not equal to the arithmetic mean of the temperatures of the top and bottom plates, i.e., $T_{\mathrm{bulk}} \neq \Delta T/2$. This is a characteristic of NOB convection, where the bulk temperature is different from the arithmetic mean of the temperatures of the top and bottom plates~\citep{Wu:PRA1991, Zhang:POF1997, Ahlers:JFM2006, Horn:JFM2013, Skrbek:JFM2015}. Higher bulk temperature in our simulation is consistent with the findings of \cite{Sameen:EPL2009} who observed that the bulk temperature rises if the thermal diffusivity increases with increasing temperature, and vice versa. {\AP Moreover, higher bulk temperature observed here is similar to that observed in fully compressible and anelastic models of convection with strong density stratification~\citep{Hurlburt:APJ1984,  Evonuk:GAFD2004}.}
We compute the horizontally- and temporally averaged vertical temperature derivatives at the plates and find that, in the convective state,
\begin{equation}
\frac{\langle (\partial T/\partial z)_{\mathrm{top}} \rangle_{A,t}}{\langle (\partial T/\partial z)_{\mathrm{bot}} \rangle_{A,t}} \approx 495, \label{eq:dtdz}
\end{equation}
which is roughly the same as for the conduction state.

Figure~\ref{fig:T_z} also shows that a boundary layer (BL) structure forms at the walls in the NOB case. The BL thicknesses are shown by vertical dashed lines. (Here and elsewhere, the thermal BL thickness is defined as the distance over which the temperature attains the center-plane value if it continued to vary linearly with the slope manifested at each wall.) Other characteristics of the NOB convection are the different thicknesses of the thermal BLs and different temperature drops within these BLs at the top and bottom plates~\citep{Wu:PRA1991}. We compute the thickness of the thermal BL $\delta_T$, using the slope method~\citep{Scheel:JFM2012}, and find that $\delta_T/H = 0.18$ and 0.0017, respectively, at the bottom and top plates. Therefore, $\delta_T$ at the bottom is roughly 100 times thicker than that at the top plate.

\begin{figure}[ht!]
\includegraphics[width=0.47\textwidth]{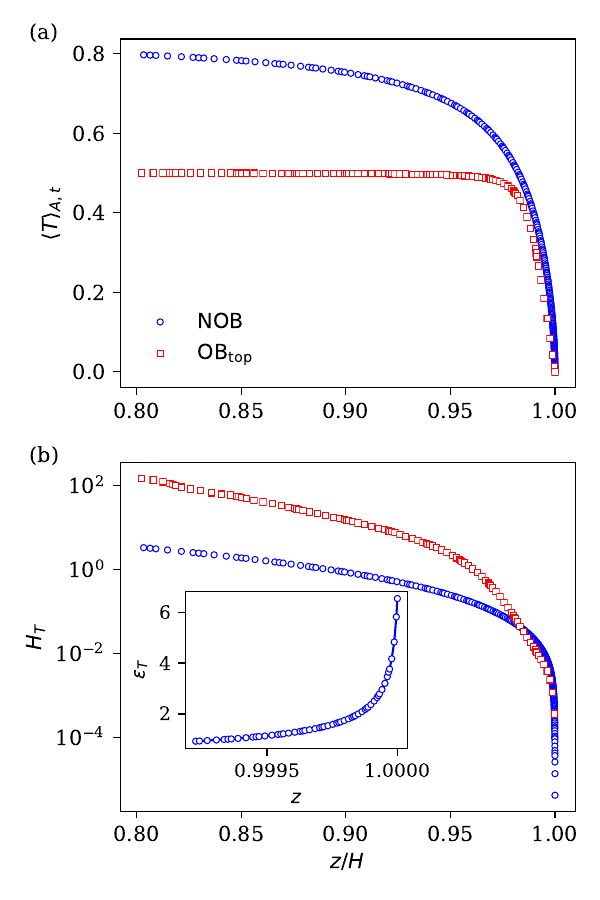}
\caption{{\AP Vertical profiles of (a) temperature and (b) temperature scale height $H_T$ near the top plate for the NOB and $\mathrm{OB_{top}}$ simulations. Inset in panel (b) shows the profile of the thermal dissipation rate for the NOB case in the vicinity of the top plate, which reveals that the profile is smooth, even though varying strongly, thus indicating the adequacy of the spatial resolution.}}
\label{fig:scale_height}
\end{figure}

{\AP A strong temperature stratification in our flow poses a great challenge to properly resolve the temperature field near the top plate. Figure~\ref{fig:T_z} shows that the temperature drop near the top boundary is similar for both the NOB and $\mathrm{OB_{top}}$ simulations. This is demonstrated more clearly by exhibiting the profiles in a magnified region near the top plate in Figure~\ref{fig:scale_height}(a). We observe that $\langle T \rangle_{A,t}(z)$ is indeed similar for the two cases in region $z > 0.98H$, and it appears that a similar resolution requirement should be obtained for the two cases near the top boundary. However, $\langle T \rangle_{A,t}$ for $\mathrm{OB_{top}}$ varies appreciably only in region $z > 0.98 H$, whereas the variation is discernible for the entire $z > 0.8H$ in the NOB case. To quantify this variation, we estimate the temperature scale height as
\begin{equation}
H_T = - \left[ \frac{ \mathrm{d} \log \langle T \rangle_{A,t} }{\mathrm{d} z} \right]^{-1} = - \frac{\langle T \rangle_{A,t}}{\mathrm{d}  \langle T \rangle_{A,t}/\mathrm{d} z} \label{eq:scale_T}
\end{equation}
and plot it as a function of $z$ in Figure~\ref{fig:scale_height}(b). We observe that $H_T$ for both the simulations are similar only in the vicinity of the top plate, and, as one moves farther from the plate, $H_T$ becomes larger for $\mathrm{OB_{top}}$ than for NOB case. This indicates that a finer resolution would be needed in the entire top region for the NOB simulation. 

Furthermore, the NOB simulation is accompanied by a strong variation of $\kappa$ near the top plate, which causes the quantities involving thermal diffusivity, such as the thermal dissipation rate $\varepsilon_T({\bm x})$ and the diffusive heat flux $J_d$, to become quite sensitive to the vertical resolution. To capture this sharp gradient in the temperature field and in the derived quantities near the top plate, we need to generate a lot more grid points there compared to that near the bottom plate.}

%===========================================================
%===========================================================

%===========================================================
%===========================================================

Thus, we have to build a highly asymmetric spectral element mesh in the vertical direction with 18, 15, and 54 elements, respectively, in the bottom, bulk, and top regions. As a consequence, the number of spectral elements for the NOB case grows by a factor of 6 at reduced polynomial order $N$ in comparison to OB cases, which demonstrates the flexibility of our numerical method while sustaining the spectral accuracy for the strongly stratified regions. To show the sufficiency of our vertical grid resolution, we display the vertical profile of the horizontally- and temporally-averaged thermal dissipation rate in the vicinity of the top plate {\AP in the inset of Figure~\ref{fig:scale_height}(b). We can see that due to very large $\partial T / \partial z$ in the vicinity of the top plate, the scalar dissipation rate, $\varepsilon_T$, becomes very large as we approach the top boundary. Nonetheless, $\varepsilon_T$ varies smoothly near the top plate in our simulation indicating that our asymmetric grid is able to capture the strong variation in the scalar dissipation rate, which is very sensitive to the spatial resolution~\citep{Scheel:NJP2013}.  Indeed, we have tested that if we use a symmetric grid as in the OB cases, strong variations of $\varepsilon_T$ and  $J_d$ cannot be captured adequately, leading to oscillations in the profiles of these quantities near the top plate.
}

In Figure~\ref{fig:T_z}, we also plot $\langle T \rangle_{A,t}(z)$ from our OB simulations, which indicate the symmetry about the midplane in the OB convection. The temperature in the bulk region is $0.5$, which is the same as the arithmetic mean $\Delta T/2$ of the top and bottom plate temperatures. The profile for the $\mathrm{OB_{top}}$ exhibits that the temperature field is well mixed in the bulk region, which lies for $0.05 \leq z \leq 0.95$. This is due to large Prandtl and Rayleigh numbers, which results in a thin thermal BLs near the top and bottom plates. The profile for the $\mathrm{OB_{bot}}$, however, exhibits a temperature gradient even in the bulk region. The thermal BLs in this case are much thicker due to smaller Prandtl and Rayleigh numbers~\citep{Scheel:JFM2016, Scheel:PRF2017, Schumacher:PRF2016, Pandey:2020a}. In OB convection, the widths of the thermal BLs at the top and bottom plates are the same, and related to the globally-averaged Nusselt number as $\delta_T/H = 1/(2 Nu)$~\citep{Chilla:EPJE2012}. We compute $\delta_T$ for the OB simulations (also indicated in Figure~\ref{fig:T_z}), and find very good agreement with those estimated from $\delta_T/H = 1/(2 Nu)$. The $\delta_T$ for $\mathrm{OB_{bot}}$ is roughly ten times thicker than that for $\mathrm{OB_{top}}$, as the Nusselt number in the former is roughly 10 times smaller than that in the latter (see Table~\ref{table:details}). 

%===========================================================
\begin{figure}[ht!]
\includegraphics[width=0.47\textwidth]{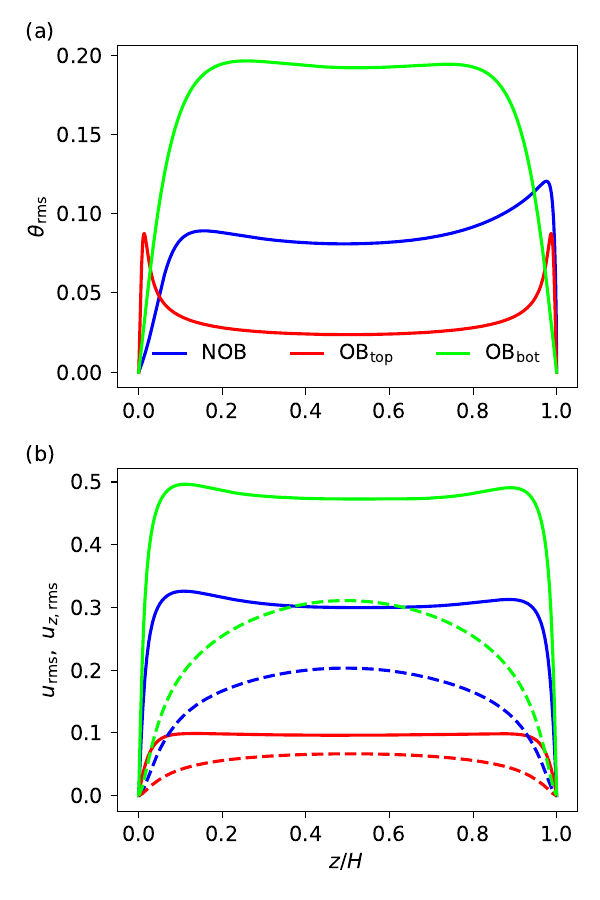}
\caption{Vertical profiles of (a) the rms temperature fluctuations from the mean temperature profile and (b) the total (solid curves) and vertical (dashed curves) rms velocities. Velocity profiles are symmetric about the midplane for all cases, whereas the $\theta_\mathrm{rms}(z)$ are symmetric only for the OB simulations.}
\label{fig:u_th_z}
\end{figure}
%===========================================================

We compute the root-mean-square (rms) temperature fluctuation $\theta_\mathrm{rms}$ defined as
\begin{equation}
\theta_\mathrm{rms}(z) = \sqrt{\langle [T - \langle T \rangle_{A,t}(z)]^2 \rangle_{A,t}},	\label{eq:theta}
\end{equation}
and show in Figure~\ref{fig:u_th_z}(a) that $\theta_\mathrm{rms}(z)$ increases with increasing distance from the horizontal plates. We observe that $\theta_\mathrm{rms}(z)$ attains a maximum near the edge of the thermal BL~\citep{Scheel:JFM2016, Pandey:2020a}, as a result of the generation and emission of thermal plumes from the thermal BL region. The thickness of the plumes is similar to the thickness of the thermal BL. The plumes originating from the bottom are hotter than ambient fluid at some distance from the  bottom plate. Thus, the decrease of $\langle T \rangle_{A,t}(z)$ with increasing distance from the bottom plate is caused primarily due the decreasing temperature of the ambient fluid as the temperature of the plumes is nearly unchanging. Therefore, the disparity between the temperatures of the plumes and the ambient fluid within the bottom BL region increases with increasing distance from the plate, which is reflected as an increasing $\theta_\mathrm{rms}(z)$ near the bottom plate. A similarly increasing contrast between the temperatures of the cold thermal plumes and the ambient fluid within the top thermal BL region causes an increase of $\theta_\mathrm{rms}$ near the top plate~\citep{Pandey:2020a}. This temperature disparity, and consequently $\theta_\mathrm{rms}(z)$, decreases outside the BL region due to an enhanced turbulent mixing of the plumes.

Figure~\ref{fig:u_th_z}(a) further exhibits that $\theta_\mathrm{rms}(z)$ is symmetric in OB convection due to the inherent up-down symmetry of the flow. For the NOB case, however, it is asymmetric; the location of the maximum of $\theta_\mathrm{rms}(z)$ is closer to the top plate than to the bottom plate because of the thinner thermal BL at the top. Furthermore, the peak of $\theta_\mathrm{rms}(z)$ is stronger near the top plate than near the bottom plate in the NOB case due to $T_\mathrm{bulk} > \Delta T/2$. As the temperature of the cold plumes within the top thermal BL is very similar to the temperature of the top plate, the disparity between the temperatures of the plumes and the ambient fluid is larger, thus causing a stronger peak in $\theta_\mathrm{rms}(z)$ near the edge of the thermal BL in the top region. {\AP The asymmetric rms temperature profile in our NOB simulation is similar to that observed in 2D anelastic convection with density stratification~\citep{Evonuk:GAFD2004}.}

We now turn to the properties of the velocity field. We compute the total and vertical rms velocity profiles defined respectively as
\begin{eqnarray}
u_{\mathrm{rms}}(z) & = & \sqrt{ \langle u_x^2 + u_y^2 + u_z^2 \rangle_{A,t} (z)} \, , \\
u_{z,\mathrm{rms}}(z) & = & \sqrt{ \langle u_z^2 \rangle_{A,t} (z)} \, ,
\end{eqnarray}
and display $u_{\mathrm{rms}}(z)$ for all the runs in Figure~\ref{fig:u_th_z}(b), which reveals that the velocity profiles are fairly symmetric about the midplane for all runs. The magnitude of $u_\mathrm{rms}$ for NOB at all depths lies between those of the OB simulations. Due to a smaller Prandtl number, the velocity magnitude is larger for $\mathrm{OB_{bot}}$ than for $\mathrm{OB_{top}}$~\citep{Schumacher:PNAS2015, Pandey:POF2016, Scheel:PRF2017}. To further compare the velocity components, we also show vertical rms velocity profiles for all the simulations in Figure~\ref{fig:u_th_z}(b), and the persistence of symmetry about the midplane; the use of temperature-dependent thermal diffusivity alone does not break the symmetry of the velocity field. We compute the viscous BL widths using the slope method~\citep{Scheel:JFM2012}, and find that, unlike different $\delta_T$, $\delta_u/H \approx 0.018$ at both the plates for the NOB simulation. 

%%%%%%%%%%%%%%%%%%%%%%=========%%%%%%%%%%
\subsection{Turbulent transport of heat and momentum}	\label{subsec:global}

\begin{figure}[ht!]
\includegraphics[width=0.47\textwidth]{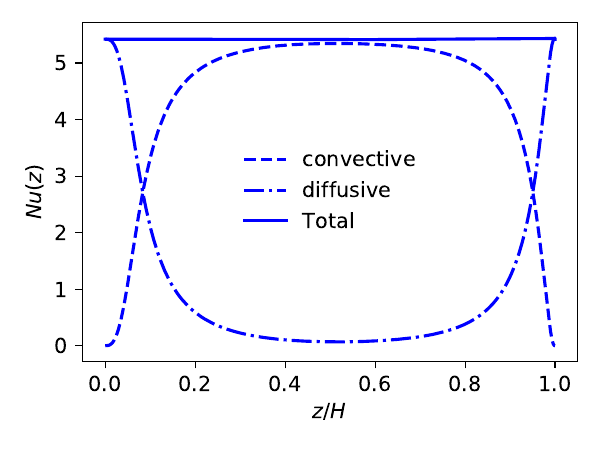}
\caption{Vertical profiles of the convective and diffusive heat fluxes for the NOB case. The total heat flux remains the same in each horizontal plane, thus exhibiting similarity with OB convection.}
\label{fig:Nu_z}
\end{figure}

It is obvious that the heat entering from the bottom plate has to go through each horizontal plane and, therefore, $Nu(z)$, defined in Eq.~(\ref{eq:Nu_z}), must be the same for each horizontal plane. To see this, we plot the convective and diffusive components of $Nu(z)$ for the NOB simulation in Figure~\ref{fig:Nu_z}, which shows that the diffusive heat flux, i.e., $- \langle \kappa(T) \partial T/ \partial z \rangle_{A,t}/J_c$, dominates the total heat transport near the plates. Figure~\ref{fig:Nu_z} also shows that the total heat flux is dominated completely by the diffusive component in the vicinity of the plates, whereas by the convective component in the region around the midplane. Interestingly, although the thicknesses of the thermal BLs are different at the top and bottom plates, the widths of the regions dominated by diffusive heat flux near the two plates are similar. This may be due to the similar thicknesses of viscous BLs near the two plates. For OB simulations too, we observe a similar contribution of the convective and diffusive fluxes. Thus, the heat transport profile for the case of temperature-dependent diffusivity remains similar to that in OB convection. 
%===========================================================
%===========================================================

We compute the volume- and time-averaged heat flux using Eq.~(\ref{eq:Nu_V}) and find that $Nu = 5.4$ for the NOB run. Moreover, as discussed in Section~\ref{subsec:flux}, the volume-averaged diffusive heat transport $J_d = -\langle \kappa(T) \partial T/\partial z \rangle_{V,t}$ in the convective state need not be the same as the diffusive flux $J_c$ in the conductive state. Therefore, we compute $J_d$ for each snapshot for the NOB simulation and find, nevertheless, that $J_d$ remains the same as  $J_c$. We also compute the volume- and time-averaged viscous and thermal dissipation rates and find that they satisfy the exact relations given by the Eqs.~(\ref{eq:exact_u}) and ~(\ref{eq:exact_T}).

Equation~(\ref{eq:Nu_V}) in OB convection simplifies to
\begin{equation}
Nu = 1 + \sqrt{RaPr} \, \langle u_z T \rangle_{V,t} \, , \label{eq:Nu_V_RBC}
\end{equation}
which yields $Nu = 3.65$ and $Nu = 35.8$ for the $\mathrm{OB_{bot}}$ and $\mathrm{OB_{top}}$ simulations, respectively. Moreover, $Nu$ computed using the exact relations in OB convection~\citep{Shraiman:PRA1990} match very well with those computed using $u_z T$, which also indicates the adequacy of the spatial resolution for the OB simulations.

The turbulent momentum transport is quantified using the Reynolds number, which is computed for the NOB run as
\begin{equation}
Re = u_\mathrm{rms} \sqrt{\frac{Ra_\mathrm{top}}{Pr_\mathrm{top}}}, \label{eq:Re_V}
\end{equation}
with $u_\mathrm{rms}$ as the root-mean-square velocity based on the volume- and time-averaged kinetic energy, i.e., 
\begin{equation}
u_\mathrm{rms} = \sqrt{ \langle u_x^2 + u_y^2 + u_z^2 \rangle_{V,t} } \, .
\end{equation}
We also use Eq.~(\ref{eq:Re_V}) to compute the Reynolds numbers for the OB simulations as the Rayleigh and Prandtl numbers remain the same throughout the simulation domain. The Reynolds numbers from the above formula are $Re = 1104,\, 345$, and 1736 for the NOB, $\mathrm{OB_{top}}$ and $\mathrm{OB_{bot}}$ simulations, respectively. The values of $Nu, Re,$ and $u_\mathrm{rms}$ for all the simulations are also listed in Table~\ref{table:details}. {\AP \cite{Rogers:PRE2003} performed simulations of high-$Ra$ 2D convection using anelastic approximation and found that the scalings of $Nu$ and $Re$ remain nearly similar to those observed in OB convection.}

%===========================================================
\begin{figure}[ht!]
\includegraphics[width=0.5\textwidth]{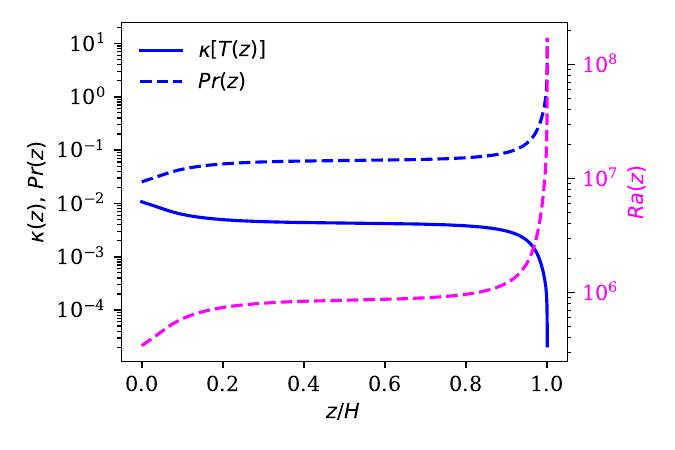}
\caption{Vertical profiles of the temperature-dependent thermal diffusivity and the resulting Prandtl and Rayleigh numbers exhibit that they vary sharply only in the region close to the top plate.}
\label{fig:kappa_z}
\end{figure}
%===========================================================

It is clear from Table~\ref{table:details} that the transport of heat and momentum as well as the rms velocity for the NOB simulation lie between the OB values, with some propensity towards the $\mathrm{OB_{bot}}$ values. To explore this, we show the vertical profile of $\kappa[T(z)]$ in Figure~\ref{fig:kappa_z}, which varies slowly in the region near the bottom plate and remains almost unvarying in the bulk region. In the vicinity of the top plate, however, it decreases very sharply to reach $\kappa_{\mathrm{top}}$, the value at the top plate. The resulting Prandtl number $Pr(z) = \nu/\kappa(z)$ and Rayleigh number $Ra(z) = \gamma g \Delta T H^3/\nu \kappa(z)$ are also plotted in Figure~\ref{fig:kappa_z}: the height-dependent Prandtl number $Pr(z)$ remains less than 0.1 over the most of convection layer and increases sharply only in the vicinity of the top plate. 

Thus, the Prandtl number in the bulk region of the NOB simulation remains approximately constant, $Pr_\mathrm{bulk} \approx 0.06$. We compute the volume- and time-averaged thermal diffusivity $\langle \kappa(T) \rangle_{V,t}$, which yields an effective Prandtl number $Pr_\mathrm{eff} = \nu/\langle \kappa(T) \rangle_{V,t} \approx 0.06$ and an effective Rayleigh number $Ra_\mathrm{eff} \approx 8 \times 10^5$. In most of the flow, these effective values are closer to the corresponding parameters of $\mathrm{OB_{bot}}$. Consequently, the global heat and momentum transports, as well as $u_\mathrm{rms}$, are comparable for the NOB and $\mathrm{OB_{bot}}$ simulations.

%%%%%%%%%%%%%%%%%%%%%%================%%%%%%%%%%%%%%%%%%%%%%%%%%%%%
\section{Flow structures}

%%%%%%%%%%%%%%%%%%%%%%=========%%%%%%%%%%
\subsection{Characteristic scales of turbulent superstructures}

\begin{figure*}[ht!]
\includegraphics[width=1\textwidth]{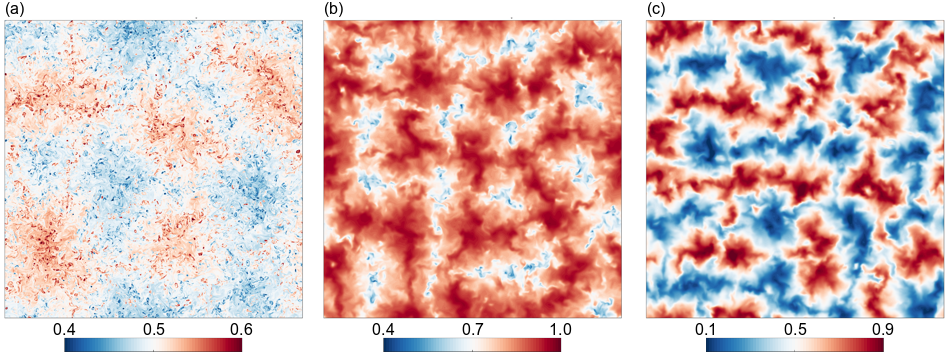}
\caption{Instantaneous temperature fields in midplane for (a) $\mathrm{OB_{top}}$, (b) NOB, and (c) $\mathrm{OB_{bot}}$ reveal turbulent superstructures. The sizes of convection structures are similar in panels (b) and (c), and are smaller than those in panel (a). The mean temperature in panel (b) is larger than 0.5, which reaffirms its essential NOB nature.}
\label{fig:tss_T}
\end{figure*}

Turbulent superstructures, whose characteristic length scales are larger than the flow depth, are observed in convection with horizontally-extended domains~\citep{Cattaneo:APJ2001,  Rincon:AA2005, BailonCuba:JFM2010, Emran:JFM2015, Pandey:Nature2018, Stevens:PRF2018, Schneide:PRF2018, Fonda:PNAS2019, Green:JFM2020, Krug:JFM2020}. Their characteristic spatial ($\lambda$) and temporal ($\tau$) scales are functions of $Pr$ and $Ra$~\citep{Pandey:Nature2018}. The superstructures in the present flows are illustrated by the instantaneous temperature field in the midplane (Figure~\ref{fig:tss_T}). The Figure exhibits alternating ridges of hot upwelling and cold downwelling fluid in the midplane. We observe that the width of the convection structures, which is half the characteristic wavelength, $\lambda_\Theta/2$, is nearly similar for $\mathrm{OB_{bot}}$ and NOB, but smaller than for $\mathrm{OB_{top}}$, consistent with \cite{Pandey:Nature2018}. Indeed, $\lambda$ increases with increasing $Ra$~\citep{Pandey:Nature2018, Stevens:PRF2018}, which results in larger $\lambda_\Theta$ for $\mathrm{OB_{top}}$.
%===========================================================
%===========================================================

%===========================================================
\begin{figure*}[ht!]
\includegraphics[width=1\textwidth]{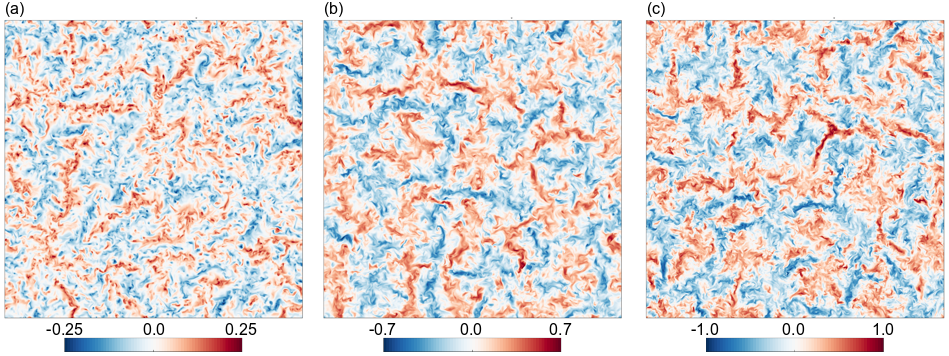}
\caption{Instantaneous vertical velocity field in midplane for (a) $\mathrm{OB_{top}}$, (b) NOB, and (c) $\mathrm{OB_{bot}}$ reveal that the characteristic lengths are similar in all cases. }
\label{fig:tss_u}
\end{figure*}
%===========================================================

The range of temperatures in Figure~\ref{fig:tss_T}(a) is smaller compared to those in other two panels. This shows that the turbulent mixing is very efficient in the bulk region for large Prandtl and Rayleigh numbers. Temperature distribution of the fluid parcels in Figure~\ref{fig:tss_T}(c) is large indicating the presence of very hot and very cold fluids in the midplane. Unlike those in panels (a) and (c) of Figure~\ref{fig:tss_T}, the temperature in panel (b) varies asymmetrically about the arithmetic mean of 0.5 for top and bottom plate temperatures. The mean temperature in panel (b) is approximately 0.82, reaffirming the NOB nature of convection in this case.

The turbulent superstructures in the velocity field, exemplified by the vertical velocity in the midplane in Figure~\ref{fig:tss_u}, show similar features for all three simulations. The magnitude of the vertical velocity is largest for $\mathrm{OB_{bot}}$ and smallest for $\mathrm{OB_{top}}$, which is also clear from the corresponding Reynolds numbers in Table~\ref{table:details}. 

Following the earlier work of~\cite{Pandey:Nature2018}, we compute the characteristic length scales of the temperature and velocity fields from the two-dimensional power spectra on a polar wavevector grid of the vertical velocity and temperature fluctuation fields in midplane, i.e., of $U(x,y) = u_z(x,y,H/2)$ and $\Theta(x,y) = \theta(x,y,z=H/2)$. Here, $\theta$ is defined as $\theta(x,y,z) = T(x,y,z) - \langle T \rangle_{A,t}(z)$. The temperature and vertical velocity power spectra are defined as~\citep{Pandey:Nature2018}
\begin{eqnarray}
E_\Theta(k) & = & \frac{1}{2\pi} \int_0^{2\pi} [\hat{\Theta}(k, \phi_k)]^2 \, d\phi_k \, , \label{eq:EThk} \\ 
E_U(k) & = & \frac{1}{2\pi} \int_0^{2\pi} [\hat{U} (k, \phi_k)]^2 \, d\phi_k \, , \label{eq:EUk}
\end{eqnarray}
where $\hat{U} (k, \phi_k)$ and $\hat{\Theta} (k, \phi_k)$ are the Fourier transforms~\citep{Verma:NJP2017} of $U(x,y)$ and $\Theta(x,y)$, respectively. The characteristic length scales are determined by finding the wavenumbers $k^{\Theta}_\mathrm{max}$ and $k^U_\mathrm{max}$ corresponding to the maximum of $E_\Theta(k)$ and $E_U(k)$, respectively, and then using $\lambda_\Theta = 2\pi/k^{\Theta}_\mathrm{max}$ and $\lambda_U = 2\pi/k^U_\mathrm{max}$.

%===========================================================
\begin{figure*}[ht!]
\includegraphics[width=1\textwidth]{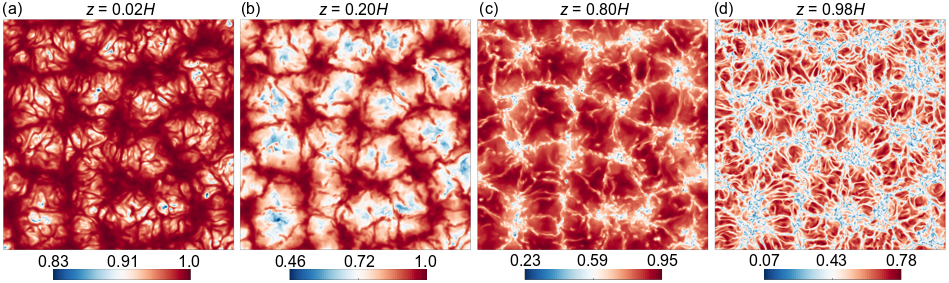}
\caption{Instantaneous temperature field in various horizontal planes for the NOB simulation. Due to decreasing thermal diffusivity with increasing altitude, temperature structures become increasingly finer. The skeleton of the pattern is the same, however, at every altitude, thus yielding approximately the same characteristic length scale over the depth. }
\label{fig:tss_T_z}
\end{figure*}
%===========================================================

We compute $\lambda_U$ and $\lambda_\Theta$ for each snapshot of our simulations and average them over many snapshots. We find that for both NOB and $\mathrm{OB_{bot}}$, $\lambda_U \approx \lambda_\Theta \approx 4H$, whereas for $\mathrm{OB_{top}}$, $\lambda_U \approx 4.7H$ and $\lambda_\Theta \approx 8H$. Thus, $\lambda_U$ is smaller than $\lambda_\Theta$ for $\mathrm{OB_{top}}$. This difference in the characteristic length scales of the temperature and velocity fields for large Prandtl number RBC is consistent with the observations of~\cite{Pandey:Nature2018} and~\cite{Fonda:PNAS2019}. We compute the characteristic temporal scale of these superstructures {\AP as~\citep{Pandey:Nature2018}
\begin{equation}
\tau(Ra, Pr) = 3 \frac{\pi( \lambda_U/4 + H/2)}{u_\mathrm{rms}}
\end{equation}
}for all the simulations and find that $\tau \approx 47 T_f, \, 30 T_f, \, 169 T_f$ for $\mathrm{NOB}, \, \mathrm{OB_{bot}}, \, \mathrm{OB_{top}}$, respectively. The characteristic scales of the turbulent superstructures are also summarized in Table~\ref{table:details}.

%%%%%%%%%%%%%%%%%%%%%%=========%%%%%%%%%%
\subsection{Depth-dependence of flow structures}

The characteristic length scales of the superstructures in OB convection are similar in the bulk and in the BLs~\citep{Pandey:Nature2018}, but the superstructures look different in the top and bottom halves for the NOB case. We show this by plotting the temperature field in horizontal cross-sections at various depths in Figure~\ref{fig:tss_T_z}. Figures~\ref{fig:tss_T_z}(a,b) show that the temperature patterns near the bottom plate are dominated by coarse (or large-scale) structures but become increasingly finer with increasing altitude. Because the thermal diffusivity decreases and the effective Prandtl number increases with altitude (given constant viscosity), the Batchelor length scale, defined as
\begin{equation}
\eta_B = (\nu \kappa^2/ \varepsilon_u ) ^{1/4}
\end{equation}
becomes smaller, reflecting the increased content of small structures in the temperature field as the altitude (i.e., the height from the bottom plate) increases. It can also be observed from Figure~\ref{fig:tss_T_z} that, even within a horizontal plane, the colder (blue) structures are finer than the hotter (red) ones. This is because the thermal diffusivity in the NOB simulation is a function of temperature, which fluctuates within a horizontal plane. {\AP We compute the rms fluctuation of $\kappa(T)$ in each horizontal plane as $\kappa_\mathrm{rms}(z) = \sqrt{ \langle [\kappa({\bm x}) - \langle \kappa \rangle_{A,t}(z)]^2 \rangle_{A,t}}$ and find that $\kappa_\mathrm{rms}(z)$ in the bulk region varies slowly and remains nearly one half of $\langle \kappa \rangle_{A,t}$ (not shown here).}

%===========================================================
\begin{figure}[ht!]
\includegraphics[width=0.47\textwidth]{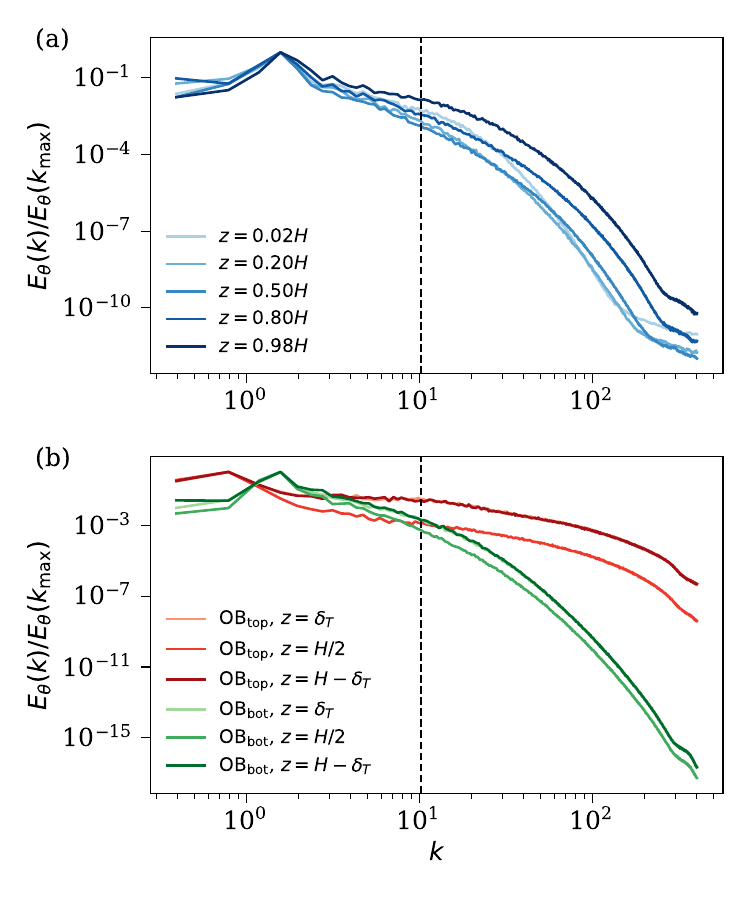}
\caption{Thermal variance spectra normalized with their maximum value for (a) NOB and (b) OB simulations. Both the panels show increasing small-scale variance when moving towards the plates. The dashed vertical line in both the panels indicate the cutoff wavenumber $k_c$ chosen to separate small and large scales.}
\label{fig:Ethk_norm}
\end{figure}
%===========================================================

To characterize these features better, we compute the power spectrum $E_\theta(k)$ defined so that its integral over the entire wavenumber range is the thermal variance on that plane, i.e., 
\begin{equation}
\int_0^{k_l} E_\theta (k) dk = \langle \theta^2(x,y) \rangle_{A} \, , \label{eq:Eth_normaliztion}
\end{equation}
where $k_l$ is the largest wavenumber in the Fourier transform. We compute $E_\theta(k)$ for an instantaneous snapshot of the NOB simulation in each horizontal plane and plot after scaling with the peak value of $E_\theta(k)$ (Figure~\ref{fig:Ethk_norm}(a)). The peak value occurs at exactly the same wavelength at each altitude, confirming that the characteristic temperature length scale remains unchanged. 
Figure~\ref{fig:Ethk_norm}(a) confirms that the total temperature variance contained in the larger wavenumbers is not a symmetric function of the distance from the midplane. In contrast, for OB simulations, the total temperature variance in the larger wavenumbers is smallest in the midplane and increases symmetrically towards either plate (Figure~\ref{fig:Ethk_norm}(b)). Moreover, due to larger $Ra$ and $Pr$, the relative spectral content at small scales is larger in $\mathrm{OB_{top}}$.
%===========================================================
\begin{figure}[ht!]
\includegraphics[width=0.47\textwidth]{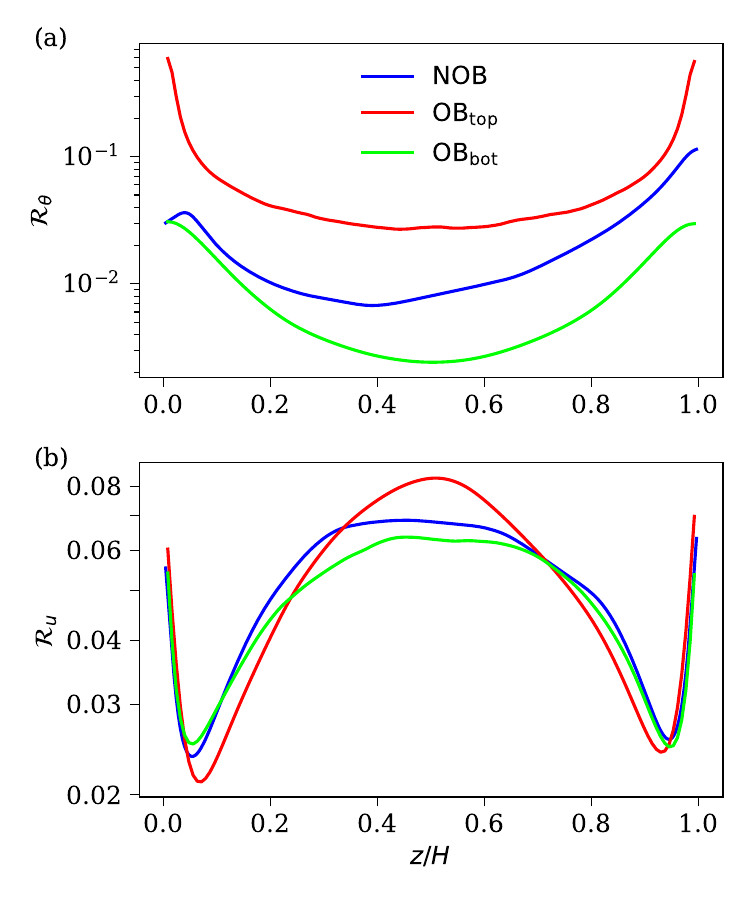}
\caption{Ratio of (a) the thermal variances and (b) the kinetic energies at the small and large scales. For the OB simulations both the $\mathcal{R}_\theta$ and $\mathcal{R}_u$ are symmetric with respect to midplane, whereas this symmetry is only observed in $\mathcal{R}_u$ for the NOB simulation.}
\label{fig:ratio}
\end{figure}
%===========================================================

To quantify the increase of small-scale structures in the temperature field with varying altitude, we compute the ratio of the total thermal variance contained in the smaller and larger scales as
\begin{equation}
\mathcal{R}_\theta = \frac{ \int_{k_c}^{k_l} E_\theta (k) dk } {\int_0^{k_c} E_\theta (k) dk },
\end{equation}
where $k_c$ is some cutoff wavenumber chosen to differentiate the small-scale structures from the larger ones. We chose $k_c = 10$, corresponding to a length scale $\ell_c/H = 2\pi/k_c = 0.62$, which is roughly half the width of the convection layer. We compute the ratio $\mathcal{R}_\theta$ in each horizontal plane and plot it in Figure~\ref{fig:ratio}(a). For all runs, $\mathcal{R}_\theta$ is smaller in the bulk and increases as the boundaries are approached. For OB simulations, the minimum of $\mathcal{R}_\theta$ occurs near the midplane, with symmetric increase in $\mathcal{R}_\theta$ towards the plates. For the NOB simulation, the minimum occurs towards the bottom plate, with asymmetric increase in $\mathcal{R}_\theta$ as the boundaries are approached; $\mathcal{R}_\theta$ increases by more than an order of magnitude from its minimum as the top boundary is approached, whereas this increase is only four-fold towards the bottom boundary.

Figure~\ref{fig:ratio}(a) also reveals that $\mathcal{R}_\theta$ is larger for $\mathrm{OB_{top}}$ at all altitudes than those for NOB and $\mathrm{OB_{bot}}$. This occurs because thermal plumes are thinner for the former case (larger $Pr$ and $Ra$), and the presence of thinner plumes everywhere contributes to the enhanced variance in the small-scale structures. Perhaps not surprisingly, the $\mathcal{R}_\theta$ curve for NOB is closer to that for $\mathrm{OB_{bot}}$ ($\mathrm{OB_{top}}$) near the bottom (top) plate. We have verified that slightly different choices of $k_c$ do not qualitatively alter the conclusions of Figure~\ref{fig:ratio}.
%===========================================================
\begin{figure*}[ht!]
\includegraphics[width=1\textwidth]{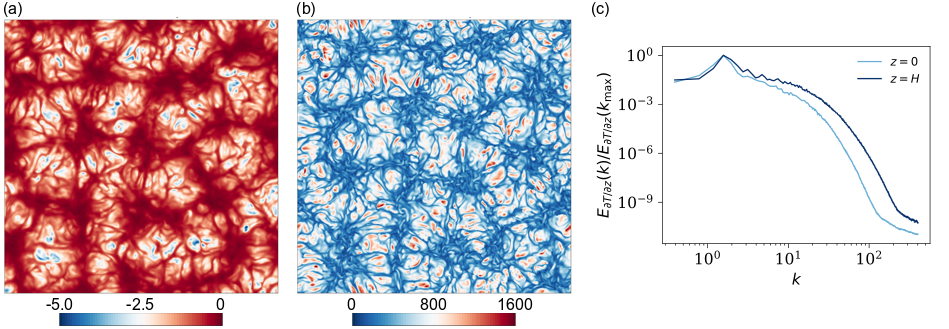}
\caption{Contours of the vertical temperature derivative $\partial T/\partial z$ at (a) the bottom and (b) the top plates show the enhanced small-scale thermal structures in the top region for the NOB simulation. Panel (c) shows the normalized power spectra of $\partial T/\partial z$ at the two plates.}
\label{fig:dtdz}
\end{figure*}
%===========================================================

The structure of the thermal BL can be explored via the vertical temperature gradient at the top and bottom plates, as it provides a blueprint of the prominent plume scales~\citep{Shishkina:JFM2006, Pandey:Nature2018}. The contours of $\partial T/ \partial z$ at the top and bottom plates for the NOB case, shown in Figure~\ref{fig:dtdz}, reveal that the patterns in $(\partial T/\partial z)_{z=0}$ (Figure~\ref{fig:dtdz}(a)) are quite similar to those in the temperature field of Figure~\ref{fig:tss_T_z}(a), but those in $(\partial T/\partial z)_{z=H}$ are different from the patterns in the temperature field even at $z = 0.98H$ (Figure~\ref{fig:tss_T_z}(d)), because of sharp variation in the temperature field near the top plate. Moreover, $\partial T/\partial z$ at the top plate contains finer structures than that at the bottom plate, despite the similarity in the skeleton of their large structures. This is in contrast to OB convection, where both fields appear very similar at the two plates~\citep{Pandey:Nature2018}. Also, $(\partial T/\partial z)_{z=H}$ is much larger than $(\partial T/\partial z)_{z=0}$, consistent with Eq.~(\ref{eq:dtdz}).

We compute the spectra of $\partial T/\partial z$ at both the plates, and show them in normalized form, $E_{\partial T/\partial z}(k)/E_{\partial T/\partial z}(k_\mathrm{max})$, in Figure~\ref{fig:dtdz}(c). The characteristic length scale, which corresponds to $k_\mathrm{max}$, is the same for both spectra, so the Figure clearly reveals the dominance of small-scale thermal structures at the top plate, consistent with the smaller thermal diffusivity towards the top.

\begin{figure}[ht!]
\includegraphics[width=0.47\textwidth]{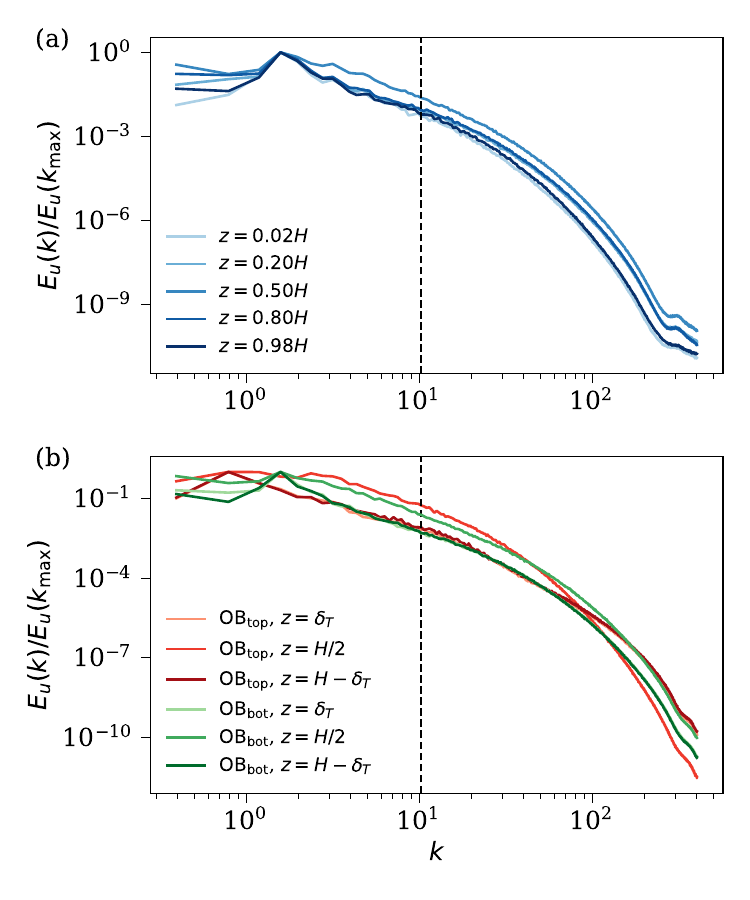}
\caption{Kinetic energy spectra normalized with their maximum value for (a) NOB and (b) OB simulations. $E_u(k)$ in both panels exhibit essential symmetry about the midplane, unlike asymmetric $E_\theta(k)$ for NOB. The dashed vertical line in both panels indicates the cutoff wavenumber $k_c$ chosen to separate small and large scales.}
\label{fig:Euk_norm}
\end{figure}

Finally, we compute the kinetic energy spectrum $E_u(k)$ in each horizontal plane as
\begin{equation}
%E_u(k) = \frac{1}{2\pi} \int_0^{2\pi} [\hat{u}_x (k, \phi_k)]^2 + [\hat{u}_y (k, \phi_k)]^2 +  [\hat{u}_y (k, \phi_k)]^2 \, d\phi_k \, \label{eq:Euk} .
E_u(k) = \frac{1}{2\pi} \int_0^{2\pi} [\hat{u}_x^2 (k, \phi_k) + \hat{u}_y^2 (k, \phi_k) +  \hat{u}_z^2 (k, \phi_k)] \, d\phi_k .\label{eq:Euk}
\end{equation}
Here $\hat{u}_i(k, \phi_k)$ is the two-dimensional power spectrum of $u_i(x,y)$ on a polar wavevector grid. We exhibit the normalized kinetic energy spectra $E_u(k)/E_u(k_\mathrm{max})$ for a few horizontal planes in Figure~\ref{fig:Euk_norm}. 
 %===========================================================
Figure~\ref{fig:Euk_norm}(a) reveals that the kinetic energy spectra for the NOB case are symmetric about the midplane, in contrast to the thermal variance spectra which exhibit increasingly finer structures when one moves upwards. Energy spectra for OB simulations in Figure~\ref{fig:Euk_norm}(b) also exhibit symmetry with respect to the midplane, similar to the thermal variance spectra.

We also compute the ratio of kinetic energy contained in small- and large-scale structures for all the simulations as 
\begin{equation}
\mathcal{R}_u = \frac{ \int_{k_c}^{k_l} E_u (k) dk } {\int_0^{k_c} E_u (k) dk },
\end{equation}
and plot them in Figure~\ref{fig:ratio}(b), which exhibits that $\mathcal{R}_u$ is symmetric about the midplane for all the simulations.  

%%%%%%%%%%%%%%%%%%%%%%================%%%%%%%%%%%%%%%%%%%%%%%%%%%%%
\section{Discussions and Conclusions} \label{sec:conclusion}

In this paper, we studied non-Oberbeck-Boussinesq convection (NOB case) by incorporating a temperature-dependent thermal diffusivity $\kappa(T)$ in the governing equations. This is important because the natural convection, such as that in the Sun, usually occur with varying molecular transport properties {\AP and accompanied by a strong stratification, and are thus non-Boussinesq~\citep{Schumacher:RMP2020}. Clearly, our model does not capture these strong variations of density and pressure. Indeed, the effects of density stratification have been investigated earlier in two- and three-dimensional convection models: The flow structure expands (compresses) the hot rising (cold sinking) plumes~\citep{Rogers:PRE2003}, thus yielding stronger downflows and weaker upflows~\citep{Brandenburg:APJ2016, Anders:APJ2019} and significant heat due to viscous dissipation~\citep{Currie:APJL2017}. 
We note that estimates in \cite{Freytag:JCP2012} or~\cite{Schumacher:RMP2020} show also that a free-fall velocity $U_f$ remains by orders of magnitude below the speed of sound $c_s$ across the solar convection zone and reaches that magnitude only at the surface which provides a further motivation to start on this significantly simpler setup.}

We chose $\kappa(T)$ as a polynomial in $T$ such that the diffusivity increases with increasing depth, whereas the kinematic viscosity was maintained constant. As a result, the Prandtl and Rayleigh numbers are different in the top and bottom regions. We studied this NOB convection using DNS in a rectangular domain of dimensions $L_x=L_y=16H$ for a $\kappa(T)$ such that $Ra$ and $Pr$ at the bottom of the flow are 500 times smaller than those at the top. For reference, we also performed two simulations of OB convection with $Ra$ and $Pr$ fixed at the top and bottom plates of the NOB case.

The use of $\kappa(T)$ in the governing equations breaks the symmetry of the temperature field about the midplane. We found that the bulk temperature is approximately 0.82, which is much higher than the arithmetic mean 0.5 of the top and bottom plate temperatures~\citep{Wu:PRA1991, Ahlers:JFM2006, Skrbek:JFM2015}. The thermal BL at the top plate is much thinner than that at the bottom plate, which makes it challenging to properly resolve the temperature field close to the top plate. Due to decreasing $\kappa(T)$, the flow structures in the temperature field become increasingly finer with increasing altitude, displaying smaller convective granules inside bigger ones. We quantified this increase by computing the ratio $\mathcal{R}_\theta$ of the total thermal variances in the small and large structures, and found that $\mathcal{R}_\theta$ increases more towards the top plate than towards the bottom.

We remark that the outcomes of the present work are qualitatively insensitive to the specific form of $\kappa(T)$ chosen here. The conclusions from this work are generally valid even if slightly different polynomial orders of $\kappa(T)$ are considered.

The symmetry of the velocity field about midplane, however, is nearly preserved in the NOB case. The rms velocity profiles and the kinetic energy spectra change symmetrically in the top and bottom region as we move away from the midplane. The profiles of convective and diffusive heat fluxes also exhibit symmetry about the midplane. Thus, the properties of heat transport and velocity field in the NOB simulation remain similar to those in OB convection. This feature may have an important implication for modeling NOB convection.

We found that the characteristic scales of the turbulent superstructures as well as the Nusselt and Reynolds numbers lie between those of the OB simulations, and are closer to the corresponding quantities for the $\mathrm{OB_{bot}}$ simulation. This may be because the volume-averaged Prandtl and Rayleigh numbers in the NOB simulation remain closer to those of the $\mathrm{OB_{bot}}$ simulation. Moreover, the characteristic length scale of the temperature field in the NOB simulation remains the same throughout the depth, despite the growing small-scale content with increasing altitude. 

Finally, we stress once more that the present study should be considered as a first step towards more realistic simulations of convection processes in sun-like stars. Our emphasis here has been on breaking the top-down symmetry at low Prandtl numbers that is characteristic of stellar cases. A further increase of the typically strong stratification should be possible by addition of a temperature dependence of the kinematic viscosity, $\nu(T)$. An additional switch to flux boundary conditions could be a further important step to understand the hierarchy of convection structures which was emphasized e.g. by \citet{Cossette:APJL2016}. Our direct numerical simulation data can also be used as a test bed for small-scale eddy viscosity and diffusivity models which will allow us to discuss the dependence of a turbulent effective Prandtl number on the molecular Prandtl number. These studies are currently underway and will be reported elsewhere.  
%%%%%%%%%%%%%%%%%%%%%%%%%%%%%%%%%%%%%%%%%%%%%%%%%%%%%%%%%%%%%%%%%%%%%%%%%%%%%%%%%%%%%%%%

\vspace{4mm}
{\AP We thank Mark Miesch for many helpful discussions on the dynamics of solar convection.} A.P. acknowledges support from the Deutsche Forschungsgemeinschaft within the Priority Programme ``Turbulent Superstructures'' under Grant DFG-SPP 1881. Computing resources at the Leibniz Rechenzentrum Garching (Germany) within the Large Scale Project pn68ni of the Gauss Centre for Supercomputing are acknowledged. This work was also supported by NYUAD Institute Grant G1502 ``NYUAD Center for Space Science."

\bibliographystyle{aasjournal}

\end{document}